\documentstyle[bobdefs,aasj,epsfig]{mn}
\newcommand{\trec}{\mbox{$\tau_{\rm rec}$}}
\newcommand{\tth}{\mbox{$\tau_{\rm th}$}}
 \newcommand{\iv}{intensity
variability} \newcommand{\noneq}{non-equilibrium}
\newcommand{\Mdot}{\langle\dot{M}\rangle}
\newcommand{\tmdot}{$\langle\dot{M}\rangle$} \newcommand{\llqq}{$L_q$}

\newcommand{\sovast}{Soviet Ast.}

\def\lesssim{\mathrel{\hbox{\rlap{\hbox{\lower4pt\hbox{$\sim$}}}\hbox{$<$}}}}
\def\gtrsim{\mathrel{\hbox{\rlap{\hbox{\lower4pt\hbox{$\sim$}}}\hbox{$>$}}}}

\makeatletter
\def\cite{\let\@citeleft(\let\@citeright)%
    \@ifstar{\citeyear}{\citefull}}
\def\citenp{\let\@citeleft\relax\let\@citeright\relax
    \@ifstar{\citeyear}{\citefull}}
\def\citefull{\def\astroncite##1##2{##1~##2}\@internalcite}
\def\citeyear{\def\astroncite##1##2{##2}\@internalcite}

\def\@citex[#1]#2{\if@filesw\immediate\write\@auxout{\string\citation{#2}}\fi
  \def\@citea{}\@cite{\@for\@citeb:=#2\do
    {\@citea\def\@citea{; }\@ifundefined
       {b@\@citeb}{{\bf ?}\@warning
       {Citation `\@citeb' on page \thepage \space undefined}}%
{\csname b@\@citeb\endcsname}}}{#1}}
\def\@cite#1#2{\@citeleft#1\if@tempswa , #2\fi\@citeright}
\def\@biblabel#1{}
\makeatother

\begin{document}

\title[Time-Variable Emission from Transiently Accreting Neutron
Stars]{Time-Variable Emission from Transiently Accreting Neutron Stars
In Quiescence due to Deep Crustal Heating}
\author[Greg Ushomirsky and Robert E. Rutledge]{Greg Ushomirsky$^1$
and Robert E. Rutledge$^2$ \\
$^1$Theoretical Astrophysics, California Institute of Technology, MS
130-33, Pasadena, CA 91125; gregus@tapir.caltech.edu;\\
$^2$Space Radiation Laboratory, California Institute of Technology, MS
220-47, Pasadena, CA 91125; rutledge@srl.caltech.edu}

\date{ 9 January 2001}

\maketitle

\begin{abstract}
Transiently accreting neutron stars in quiescence ($L_X$\approxlt
\ee{34} \cgslum) have been observed to vary in intensity by factors of
few, over timescales of days to years.  If the quiescent luminosity is
powered by a hot NS core, the core cooling timescale is much longer
than the recurrence time, and cannot explain the observed, more rapid
variability.  However, the non-equilibrium reactions which occur in
the crust during outbursts deposit energy in iso-density shells, from
which the thermal diffusion timescale to the photosphere is days to
years.  The predicted magnitude of variability is too low to explain
the observed variability unless -- as is widely believed -- the
neutrons beyond the neutron-drip density are superfluid.  Even then,
variability due to this mechanism in models with standard core
neutrino cooling processes is less than 50 per cent -- still too low
to explain the reported variability.  However, models with rapid core
neutrino cooling can produce variability by a factor as great as 20,
on timescales of days to years following an outburst. Thus, the
factors of $\sim$few intensity variability observed from transiently
accreting neutron stars can be accounted for by this mechanism only if
rapid core cooling processes are active.

\end{abstract}

\begin{keywords}
stars: neutron -- X-rays: binaries -- nuclear reactions,
nucleosynthesis, abundances
\end{keywords}

\section{Introduction} 

\label{sec:intro}

Many neutron star (NS) X-ray binaries go through accretion outbursts
(\lx$\sim$\ee{37} \cgslum) followed by long periods (months--decades) of
relative quiescence (\lx\approxlt\ee{34} \cgslum).  The origin of
these outbursts remains under debate, although many agree that in the
wider binaries an accretion disk instability is the cause
\cite{jvp96}.  During these outbursts, on the order of $\sim$\ee{23}~g
may be accreted, over a period of $\sim$few-30 days (for a review of
these transients see \citenp{tanaka96,campana97,chen97}).

Following an outburst, the NSs return to quiescence, and have
typically been detected with luminosities in the \ee{32}--\ee{33}
\cgslum\ range
\cite{jvp87,verbunt94,asai96a,asai96b,campana98b,garcia99,rutledge99,campana00,rutledge00}.
The quiescent X-ray spectrum has been described as being comprised of
two components, a soft (black-body $kT\sim$0.2 keV) thermal component,
and  a power-law component dominating the emission above 2 keV
\cite{asai96b,campana98a,campana00}.

The basal luminosity has been attributed to continued accretion in
quiescence \cite{jvp87}, possibly through an advection-dominated
accretion flow \cite{narayan97a,menou99}; accretion onto the NS
magnetosphere \cite{campana97} following a ``propeller phase'', in
which the NS magnetosphere is larger than the keplerian orbit with an
orbital period equal to the spin period of the NS
\cite{ill75,stella86}; and to thermal emission from a hot NS core,
heated by non-equilibrium reactions deep in the NS crust \cite[BBR98
hereafter]{brown98}. Of these possibilities, only deep-crustal heating
predicts the similar luminosities observed in the thermal spectral
component of these systems. The observed temperatures of the thermal
component ($kT$=0.08--0.20 keV) are as expected in this scenario.  In
addition, the emission area radii of the thermal component are
consistent with theoretically predicted NS radii ($\sim$10 km;
\citenp{rutledge99,rutledge00}); this supports the interpretation of
this spectral component as a thermal NS photosphere.  It is possible
that the quiescent luminosity is due to a combination of these three
mechanisms, in which case they are only separable by their spectral
and intensity variability properties.

The detected flux during quiescence from some of these transients has
been observed to vary, by a factor of as much as $\sim$3 or more, over
timescales of days to years.  No clear cause of this variability has yet
emerged.  There are three means discussed in the literature:

\begin{enumerate}
\item {\em Variable accretion onto the NS surface.}  If accretion onto
the NS surface dominates the luminosity in quiescence, changes in the accretion rate
could account for variations in the thermal (\kteff$\sim$0.2 keV) part
of the spectrum, since the photospheric spectrum at the low implied
accretion rate is thermal in the absence of a shock \cite{zampieri95},
assuming quasi-steady-state accretion at the appropriate accretion
rates. 
\item {\em   Variable accretion onto the NS
magnetosphere.}  The power-law spectral ``tail'' which can dominate
emission above 2 keV may be due to accretion onto a NS magnetosphere
\cite{campana98b}.  Variations in accretion rate onto the magnetosphere
would cause \iv\ in this power-law component. 
\item {\em  Variable
absorption column density (\nh).}  Outbursts may be accompanied by
outflows, increasing locally the column density of absorbing material.
This would cause variations in the absorption column density (\nh),
which may account for the factors of 2-3 variation in intensity
observed \cite{rutledge99}.  
\end{enumerate}

\noindent Moreover, \iv\ due to accretion -- either photospheric or
magnetospheric -- would likely be stochastic; there have been no
theoretical estimations of its magnitude or variability timescale.  In
the variable-\nh\ scenario, the amount of absorption should decrease
(and intensity, increase) with time after outburst.

In this paper, we investigate the time dependence of the thermal
emission mechanism proposed by BBR98. Compression of the NS crust by
accretion during outbursts induces electron captures, neutron
emissions, and pycnonuclear reactions deep in the crust (at densities
$>10^9$~g~cm$^{-3}$), and these reactions deposit $Q_{\rm
nuc}\approx1$~MeV per accreted baryon of heat into the crust
\cite{haensel90}. BBR98 argued that these reactions heat the NS core
to an equilibrium temperature $\approx10^8$~K, and, during quiescence,
the hot core shines with a typical luminosity $L_q\approx\Mdot(Q_{\rm
nuc}/m_b)$, where $\Mdot$ is the mean accretion rate, averaged over
the thermal time of the core, i.e., many recurrence intervals \trec.
Consequently, the core luminosity is not expected to change on
timescales shorter than its thermal time, i.e., $\sim10^5$~yrs.

{\it Transient\/} X-ray emission from the \noneq\ reactions has not
been previously considered.  Similar work regarding transient energy
deposition in NSs has either focussed on the atmosphere, well above
the crust \cite{eichler89}, or investigated the thermal response to
pulsar glitches \cite{vanriper91,chong94,hirano97,cheng98}.  However,
the crucial difference is that the depths and amounts of energy
deposited by \noneq\ reactions can, at least in principle, be obtained
from an {\it ab initio\/} calculation (and have been, for the case of
iron, by \citenp{haensel90}), while, on the other hand, the total
amount of energy deposited by the crust-breaking glitches is not
precisely known (it is bounded by $2 I\Omega\delta\Omega$), and
neither the depth nor the distribution of energy deposition is well
constrained.

Colpi \etal\ \cite*{colpi00b} examined the effect of \noneq\ reactions
on the temperature of the core.  They found that in transient NSs with
recurrence timescales of $\sim$1 year, the core heats to an
equilibrium temperature in $\sim$\ee{4} yrs, after which the core
temperature varies by $\sim$0.5\% in response to the periodic input of
energy from the \noneq\ reactions during individual accretion
outbursts.  They confirmed the steady-state assumption made by BBR98.
Our study is complimentary to that of Colpi \etal, as we examine the
time dependence of the temperature in the crust and the thermal
luminosity from the NS surface between outbursts.  We follow in detail
the thermal relaxation of the crust and find that, depending on the
microphysics and the accretion history, the magnitude of the
variability can be as small as $<1$\%, or as large as a factor of 20.
Only the reactions that deposit energy at depths where
\tth$\lesssim$\trec\ lead to variable thermal emission.  The
luminosity due to individual reactions is largely blended together,
simply because the difference between the \tth\ of adjacent reactions
is typically much shorter than \tth.  However, the emission from the
reactions in the outer crust as a whole is typically well-separated
from that due to the reactions in the inner crust, resulting in
characteristic ``double-hump'' luminosity evolution.  In addition, the
amplitude of the variability is not simply related to the energy
deposited in the individual reactions.  We quantify the dependence of
this variability on the conductivity of the crust and the physics of
neutrino emission in the core.

In \S\ref{sec:vari} we briefly recount the observational evidence for
quiescent intensity variability.  In \S\ref{sec:calc}, we describe
detailed calculations of time evolution in the thermal NS flux, in
response to a ``delta-function'' (1-day long) accretion event, for
different time-averaged accretion rates (and thus core temperatures),
different dominant conductivies and core neutrino cooling
prescriptions We also present the resulting light curves in this
section.  Finally, in \S\ref{sec:discuss}, we compare these results
with observations, and in \S\ref{sec:conclude}, we conclude.

\section{Observations of Variability in Quiescence}
\label{sec:vari}

Intensity variability in quiescent transients is, at present, not
well-studied observationally, largely due to the low signal-to-noise
of the data, from a small number of observations (only a few of which
are made with the same instrumentation), and different assumed
intrinsic energy spectra.  Multiple observations in quiescence (at
\llqq\approxlt \ee{34} \cgslum) are found in the literature only for
three NSs, \cenx4, 4U~2129+37, and \aql.

For Cen~X-4, Van Paradijs \etal\ \cite*{jvp87} found the luminosity
increased by a factor of $\sim 2-5$ over $\sim$5.5 yr (Jul 1980--Feb
1986), for the same assumed thermal bremmstrahlung spectrum.  There
was no intervening outburst observed, and the observations were made
using two different instruments ({\em EXOSAT}/LE and {\em
Einstein}/IPC).  Campana \etal\ \cite*{campana97} reanalysed these
data, and contrarily concluded that they are consistent with the same
luminosity.  Also, Campana found that in observations with {\em
ROSAT}/HRI of \cenx4\ over a 4-8 day period, the source countrate
varied by $\sim$3, with an average luminosity of $\sim$7\tee{31}
\cgslum; Campana did not discuss whether the 1995 {\em ROSAT}/HRI
observations were consistent with the luminosities of the {\em EXOSAT}
and {\em Einstein} observations; a comparison between their values and
passbands indicates that the luminosities of Van Paradijs \etal\ were
more than a factor $3$ greater, although part of this may be due to a
different assumed spectrum.  Finally, Rutledge \etal\
\cite*{rutledge01b} found that Cen~X-4 varied by $<$18\% (0.2-2 keV)
during a 10 ksec observation, while the 0.5-10.0 keV luminosity had
decreased by 40\ppm8\% in comparison with an \asca\ observation taken
5 years previously.

Aql~X-1 was also found to have a variable flux in quiescence
\cite{rutledge99} between three observations with \asca\ (Oct 1992,
Mar 1993, and Oct 1996). If the same intrinsic spectrum is assumed,
the three observations are inconsistent with having the same intensity
with high confidence (probability=\ee{-6}). The difference in observed
flux is a factor of $\sim$2-3.

4U~2129+37 decreased in flux by a factor of 3.4\ppm0.6 between Nov-Dec
1992 ({\em ROSAT}/HRI) and March 1994 ({\em ROSAT}/PSPC)
\cite{garcia99,rutledge00} in the unabsorbed luminosity, for the same
assumed spectrum.

The observations of Cen~X-4, with {\em ROSAT}/HRI, and of Aql~X-1,
with \asca\ are the only two instances of repeated observations with
the same instrumentation.  Cen~X-4 has not been observed in outburst
since May, 1979 \cite{chen97}, while Aql X-1 goes into outburst every
$\sim$220 days, and it therefore seems likely that there were
intervening outbursts between the three \asca\ observations.  

Thus, observations of transiently accreting NSs in quiescence, taken
at face value, indicate that their quiescent luminosities vary by a
factor of up to 3$-$5 on timescales of days to years.  However, it is
not clear what fraction of this variability is intrinsic, and what can
be attributed to systematic differences in instrumentation.

\section{Thermal Emission from the Crust}
\label{sec:calc}

We now describe our simulations of the thermal relaxation of the crust
following an accretion outburst, and the resulting time dependence of
the quiescent thermal emission.  As argued by BBR98 and confirmed by
Colpi \etal\ \cite*{colpi00b}, the nuclear energy release in the crust
heats the NS core to a temperature corresponding to steady accretion
at the corresponding time-averaged accretion rate.  We performed
several simulations starting with a very cold NS and subjecting it to
a series of accretion outbursts and found that indeed, the core is
heated to the appropriate temperature on a \ee{4}~yr timescale, in
agreement with Colpi \etal~\cite*{colpi00b}.  For the remainder of our
analysis we first construct a steady-state thermal model corresponding
to the time-averaged $\langle\dot{M}\rangle$, and then subject it to
outbursts. After several outbursts, the model reaches a limit cycle.
In this paper, we report the results of only these latter simulations.
Our accretion events (``outbursts'') are 1-day long
``delta-functions'', with no accretion outside of these events. More
realistic time-evolved light-curves may in principle be found by
convolving our ``delta-function'' thermal response light curves with a
more realistic outburst accretion time-dependent profile for the small
perturbation light curves ($\delta L/L$\approxlt 1); lightcurves in
which the variations are larger would require a more detailed
calculation.  We simulate two different outburst recurrence
timescales (1 yr and 30 yr) and three time-averaged accretion rates
(\ee{-10}, \ee{-11}, and \ee{-12} \msun/yr).  As we are primarily
interested in illustrating the response of the NS surface flux from
the crust and core to a ``delta-function'' accretion profile, it is
unimportant that the implied 1-day accretion rates may be
super-Eddington, and we do not include accretion luminosity in our
results.

\subsection{Microphysics of the Crust and Core}
\label{sec:microphysics}

The hydrostatic and steady-state thermal models of the crust used in
this paper are substantially identical to the ones used by Ushomirsky,
Cutler, \& Bildsten~\cite*{UCB99}.  We briefly recount the major
ingredients of these models, and then describe our time-evolution
code.

Since the thermal timescale for the crust (days to years, see below)
is so much longer than the sound crossing time (milliseconds), and the
equation of state is nearly independent of temperature, the
hydrostatic structure of the crust is effectively decoupled from its
thermal evolution.  Therefore, one can solve the thermal evolution
equations assuming that the density and pressure as a function of the
position in the star are fixed. This approach to modeling the thermal
state of a neutron star is fully described by Brown~\cite*{brown00}
and is followed here.

The composition of the crust (i.e., the mass $A$ and charge $Z$ of the
nuclei, as well as the neutron fraction $X_n$) is taken from the
tabulation of Haensel \& Zdunik \cite*{haensel90,HZ90b}.  The pressure
is the sum of contributions from degenerate, relativistic electrons
and free neutrons \cite{negele73} at densities exceeding neutron drip
($\rho_{\rm nd}\approx (4-6)\times10^{11}$~g~cm$^{-3}$,
\citenp{HZ90b}). With the equation of state as described above, we
solve the Newtonian equations of hydrostatic balance (which, for a
thin crust, are exactly equivalent to the fully relativistic
equations).  We take $M=1.4M_\odot$, and $R=10$~km.  Our crust has a
thickness of 1.1~km and a mass of $0.06M_\odot$.

With the hydrostatic structure specified, we solve the heat equation,
\begin{equation}\label{eq:heat}
\rho c_v \frac{\partial T}{\partial
t}=\frac{1}{r^2}\frac{\partial}{\partial r}
\left(r^2 K \frac{\partial T}{\partial r}\right)+
\rho(\epsilon_{\rm nuc}-\epsilon_\nu),
\end{equation}
where $c_v$ is the heat capacity per gram (which has contributions
from the ionic lattice, degenerate electrons, and free neutrons in the
inner crust), $\epsilon_{\rm nuc}$ is the nuclear energy release,
$\epsilon_\nu$ is the neutrino emissivity, and $K$ is the thermal
conductivity. We neglect the overall downward motion of the material
due to accretion, as well as the heat release due to compression
(these effects are not important in the crust).  Therefore, the only
energy source in the crust is the heat release due to non-equilibrium
electron captures, neutron emissions, and pycnonuclear reactions.  The
detailed treatment of the energy release is described in Ushomirsky
\etal\ \cite*{UCB99}; we note here that $\epsilon_{\rm nuc}$ is
proportional to the instantaneous mass accretion rate $\dot{M}$. The
total energy released in the crust is taken to be $Q_{\rm nuc}=\int
4\pi r^2 \rho\epsilon_{\rm nuc} dr= 1.45$~MeV per accreted baryon
\cite{haensel90}.  At the temperatures of interest, neutrino losses
{\it in the crust\/} are negligible.

Instead of fully modeling the thermal evolution of the NS core, we
presume that it is isothermal, and is characterised by a single
temperature $T_{\rm core}$ which is equal to the temperature of the
crust at the interface with the core, $T(r_{\rm core})$.  The core
temperature obeys
\begin{equation}\label{eq:tcore}
C_{\rm core} \frac{d T(r_{\rm core})}{dt}
=4\pi r_{\rm core}^2 K\left.\frac{\partial T}{\partial
r}\right|_{r_{\rm core}}-L_\nu,
\end{equation}
where $C_{\rm core}$ is the total heat capacity, and $L_\nu$ is the
total neutrino emissivity of the core (see below).  This approximate
treatment is justified because of the core's large thermal
conductivity and heat content.  In effect, the temperature of the core
changes by a negligible amount compared to the temperature variations
in the crust, and the core acts as an energy sink at the bottom edge
of the crust.  With this approximation, we do not need to integrate
the heat equation in the core, and instead use Eq.~(\ref{eq:tcore}) as
a boundary condition for Eq.~(\ref{eq:heat}) at the bottom edge of the
crust.

To survey the parameter space, we use two different prescriptions for
core cooling.  ``Standard'' cooling models use modified Urca
\cite{friman79,yakovlev95} and $e$-$e$ neutrino bremsstrahlung
\cite{kaminker99:ee-brem}, while ``rapid'' cooling models assume the
presence of a pion condensate \cite{shapiro83}\footnote{Instead, one
could assume that the NS mass is high enough to allow direct Urca
neutrino emission (see, e.g., Colpi \etal\ \cite*{colpi00b}), as the
emissivities are similar.}.  Nucleon superfluidity significantly
reduces neutrino emissivity in the core (see
\citenp{yakovlev99:_coolin} for an in-depth review).  In our
calculation we use superfluid parameters collated by
Brown~\cite*{brown00}.  Maxwell \cite*{maxwell79:_neutr} argued that
at temperatures well below the superfluid transition temperature
$T_c$, this suppression is exponential.  However, detailed
calculations, summarized by Yakovlev \etal\ \cite*{yakovlev99:_coolin}
show that this suppression is weaker by a power-law factor in $T/T_c$,
leading to lower core temperatures than one would expect based on the
simple exponential reduction in emissivity.

Core heat capacity is the sum of contributions from electrons,
neutrons, and protons.  However, superfluidity alters the heat
capacity of the nucleons (see \citenp{yakovlev99:_coolin} for an
in-depth review).  At $T\ll T_c$, the reduction of the heat capacity
is similar to the suppression of neutrino emissivity.  Maxwell
\cite*{maxwell79:_neutr} gives a fitting formula for $^1$S$_0$
superfluid (applicable to neutrons in NS crust and protons in the
core), while Levenfish \& Yakovlev \cite*{levenfish94} give fitting
formulae for the $^3$P$_2$ superfluid as well (applicable to neutrons
in the core).  At temperatures of interest ($\sim10^7-10^8$~K) electrons
dominate the heat capacity.

Throughout the crust, heat is transported by degenerate, relativistic
electrons, and the thermal conductivity $K$ is very sensitive to the
purity of the crust.  If the crust is a pure crystal, electron-phonon
scattering \cite{baiko95} sets the conductivity.  However, crusts of
accreting neutron stars are unlikely to be pure crystals, since they
are composed of (likely quite impure) products of nuclear burning in
the upper atmosphere.  In fact, calculations of Schatz \etal\
\cite*{Schatz99} suggest that the typical values of the impurity
parameter $Q_{\rm imp} = Y_{\rm imp}(Z_{\rm imp}-\langle Z\rangle)^2$,
where $Z_{\rm imp}$ and $Y_{\rm imp}$ are the charge and the fraction
of the impurities, could be comparable to the average charge $\langle
Z\rangle^2$.  Following Brown~\cite*{brown00}, we set the absolute
lower bound on the conductivity by using electron-ion scattering
\cite{yakovlev80:_therm}, which has the same form as electron-impurity
scattering with $Q_{\rm imp}=Z^2$. The much smaller conductivity in
the case of electron-impurity scattering with large $Q_{\rm imp}$
leads to much longer crustal thermal time, and much smaller
variability of quiescent thermal emission.  This approach allows us to
survey the entire range of possibilities for the crustal
conductivity.\footnote{The presence of impurities does not
dramatically affect the heat capacity of the ionic lattice.  While the
phonon spectrum will be affected by the presence of impurities, the
heat capacity should still approach the Debye limit.}

At the top of the crust ($\rho\approx10^8$~g~cm$^{-3}$) we utilize the
results of Potekhin \etal\ \cite*{potekhin97} for the relation between
the core temperature to the surface temperature.\footnote{Potekhin
\etal\ \cite*{potekhin97} express the surface temperature in terms of
the temperature in the isothermal region of the crust, which they
assume to be at $\rho>10^{10}$~g~cm$^{-3}$.  However, since the
authors assume that the crust has high thermal conductivity, the
temperature difference between $10^{10}$~g~cm$^{-3}$ and
$10^{8}$~g~cm$^{-3}$ is negligible in their models.}  They extended
the original calculations of Gudmunsson \etal\ \cite*{gudmundsson83}
by considering the opacities of accreted envelopes, rather than iron
envelopes.  Because of the smaller opacity of a light-element
envelope, a $\sim$50\% smaller core temperature is necessary to carry
a given flux (see also \citenp{blandford83}).  The thermal time at the
top of the crust, where the outer boundary condtion is applied, is
$\lesssim1$~day in our models, so our simulations cannot follow
variability on shorter timescales.

\begin{figure*}
\begin{center}
\epsfig{file=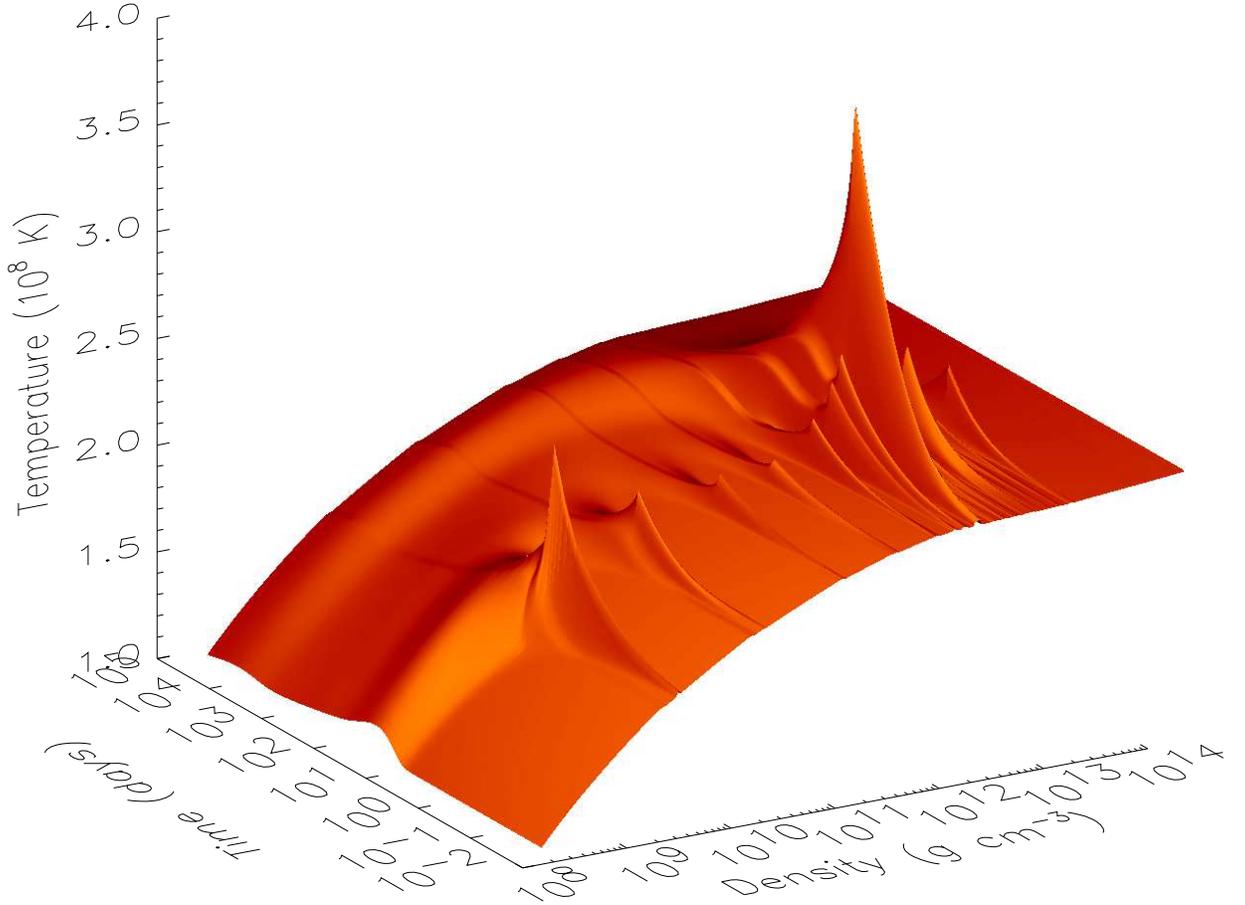}
\end{center}
\caption{\label{fig:tevol} Time evolution of the temperature of the
crust for a model with standard cooling and low thermal conductivity
crust following a 1-day outburst and $\Mdot=10^{-10}
M_\odot$~yr$^{-1}$, with a recurrence time of 30~years. The $x$ axis
is the density, covering the entire crust of the neutron star; $y$
axis is the time, and $z$ axis is the temperature. }
\end{figure*}

\subsection{Time-Averaged Quiescent Luminsity}
\label{sec:time-average}

Before describing the results of our time evolution simulations, we
discuss the thermal luminosity of neutron stars assuming steady
accretion at the rate $\Mdot$ corresponding to the time average over
the outburst recurrence interval \trec.  This provides a reasonable
estimate of the average luminosity level of neutron star transients.
The variations of the luminosity discussed in
\S~\ref{sec:time-variable} are excursions about this average level. 
As outlined in \S~\ref{sec:microphysics}, we survey the parameter
space by considering two different strengths of core neutrino emission
(standard and enhanced), and two different assumptions regarding the
conductivity of the crust. 

The average core temperature and luminosity of the NS are set by the
balance between the heat input due to \noneq\ reactions in the crust
(at the rate $Q_{\rm nuc}\langle\dot{M}\rangle/m_b$) and heat loss due
to neutrino emission from the core ($L_\nu$) and photon luminosity
from the surface ($L_\gamma$).  At the low core temperatures
characteristic of NS transients ($T_{\rm core}\lesssim10^8$~K)
modified Urca neutrino emission (standard cooling case) is
substantially suppressed by nucleon superfluidity and cannot compete
with photon losses from the surface of the star, $L_\nu\ll L_\gamma$.
Therefore, most of the heat deposited by the nonequilibrium reactions
is radiated from the surface, i.e., the quiescent luminosity is just
(BBR98)
\begin{eqnarray}
\label{eq:Lq-normal-cool}
L_q&\approx& \frac{Q_{\rm nuc}}{m_b} 
{\langle\dot{M}\rangle} \\ \nonumber
&=&8.7\times10^{33}\left(\frac{\langle\dot{M}\rangle}{10^{-10}
{\rm~}M_\odot{\rm~yr}^{-1}}\right)
\left(\frac{Q_{\rm nuc}/m_b}{1.45{\rm~MeV}}\right)
{\rm~erg~s}^{-1}.
\end{eqnarray}
This estimate is {\it independent\/} of the conductivity of the crust.
The core temperature, however, depends on the crustal conductivity. If
the crustal conductivity is large (i.e., set by electron-phonon
scattering), the crust is nearly isothermal, and the core temperature,
$T_{\rm core}\approx
1.2\times10^8$~K~$(\Mdot/10^{-10}M_\odot$~yr$^{-1})^{0.41}$, can be
obtained by simply using the relation of Potekhin \etal\
\cite*{potekhin97}.  If the crust has low thermal conductivity, then a
substantial temperature gradient is needed to carry the heat from the
inner crust to the surface, and the core temperature is $T_{\rm
core}\approx
2.2\times10^8$~K~$(\Mdot/10^{-10}M_\odot$~yr$^{-1})^{0.45}$ (obtained
by fitting our detailed calculations).

On the other hand, if rapid cooling processes are allowed in the core,
then, despite the suppression of neutrino emission by superfluidity,
most of the heat deposited by \noneq\ reactions is radiated away by
neutrinos, $L_\nu\gg L_\gamma$.  The core temperature is then set by
balancing the energy input from the crustal reactions with
neutrino-luminosity.  Except at very low accretion rates
($\lesssim5\times10^{-12} M_\odot$~yr$^{-1}$), where superfluid
suppression of neutrino emission makes $L_\gamma$ comparable to
$L_\nu$, a good fit to the results of our calculations for the
particular case of enhanced neutrino emission from a pion condensate
is $T_{\rm core}\approx
2.7\times10^{7}$~K~$(\Mdot/10^{-10}M_\odot$~yr$^{-1})^{0.09}$, {\it
regardless\/} of the conductivity of the crust.  The thermal
luminosity, however, depends on the relation between the core
temperature and that at the top of the crust.  When the conductivity
is determined by electron-phonon scattering (i.e., $K$ is large), the
crust is nearly isothermal, and, using the relation of Potekhin \etal\
\cite*{potekhin97}, we find
\begin{equation}\label{eq:Lq-enh-eph}
L_q\approx 2.2\times10^{32}
\left(\frac{\langle\dot{M}\rangle}{10^{-10}
{\rm~}M_\odot{\rm~yr}^{-1}}\right)^{0.22}
\left(\frac{Q_{\rm nuc}/m_b}{1.45{\rm~MeV}}\right)^{0.22}
{\rm~erg~s}^{-1},
\end{equation}
which agrees well with our detailed calculations.  On the other hand,
if the crustal conductivity is low, the relation between the core and
the surface temperature is nontrivial, since an appreciable
temperature gradient between neutron drip and the core is needed to
carry the flux from the \noneq\ reactions into the core.  In this
case, the crust is warmer than in the electron-phonon scattering case,
and the thermal luminosity is
\begin{equation}\label{eq:Lq-enh-ei}
L_q\approx 7\times10^{32}
\left(\frac{\langle\dot{M}\rangle}{10^{-10}
{\rm~}M_\odot{\rm~yr}^{-1}}\right)^{0.8}
\left(\frac{Q_{\rm nuc}/m_b}{1.45{\rm~MeV}}\right)^{0.8}
{\rm~erg~s}^{-1}
\end{equation}
(which we obtained by fitting the results of our detailed
calculations).  Comparing equation~(\ref{eq:Lq-normal-cool}) with
(\ref{eq:Lq-enh-eph}) and (\ref{eq:Lq-enh-ei}), we see that when
enhanced neutrino cooling processes are allowed in the NS core, only
less than 10\% of the \noneq\ energy release is radiated from the
surface, while the remaining $>90$\% of the heat is emitted in
neutrinos.  Regardless of the crustal conductivity, the core
temperature in the rapid cooling case is set entirely by the core
neutrino emission.  However, depending on the conductivity of the
crust, a temperature gradient between the core and the surface may or
may not be present, resulting in different surface temperatures and
photon luminosities (cf. Eqs.~(\ref{eq:Lq-enh-eph}) and
(\ref{eq:Lq-enh-ei})).  {\em Thus, neutron stars with rapid cooling
processes active in the core appear dimmer than those with just
standard cooling, and, when their cores are very cold due to rapid
neutrino cooling, NSs with high conductivity crusts will appear dimmer
than those with lower conductivity crusts.}

\subsection{Time-Variable Emission}
\label{sec:time-variable}

As we now describe, the approximate steady-state luminosity is only
a part of the story. Depending on the microphysics of the crust and
core, the time-dependent thermal luminosity of the neutron star may
either always be very close to the steady-state estimate ($\delta L/L<$
1\%), or vary wildly around it ($\delta L/L$\approxgt 1).  

In Figure~\ref{fig:tevol} we display the time evolution of the
temperature in the NS crust for a model with low thermal conductivity,
standard neutrino emission, $\Mdot=10^{-10} M_\odot$~yr$^{-1}$, and
\trec=30~yrs.  Other models are similar qualitatively, but, of course,
differ in the magnitude of variability.  The slices of the figure in
the $x-z$ plane represent the temperature in the crust as a function
of density, and the $y$ axis is the time (in days) since the beginning
of a 1-day outburst.  At the early times, the temperature rises
locally at the densities where the energy is deposited. It reaches the
maximum after 1 day (i.e., at the end of the outburst). In the ensuing
cooling period, the heights of the temperature peaks decrease, and
their widths increase due to heat diffusion. After $\sim6$~days the
``heat wave'' due to the very first reaction, at
$\rho=1.5\times10^{9}$~g~cm$^{-3}$ reaches the top of the crust.  This
temperature increase at the top of the crust is then directly
translated into the increase in the thermal luminosity of the NS
\cite{potekhin97}.  At much later time ($\sim1000$~days after the
outburst) the heat wave due to the reactions in the inner crust, at
$\rho\gtrsim10^{12}$~g~cm$^{-3}$ reaches the surface. Even though the
energy release in these reactions is much larger, the change in the
outer boundary temperature and, hence, the variation of the thermal
luminosity is much smaller, because the temperature peak spreads out
to a width corresponding to roughly the thermal time at the deposition
depth as it travels towards the surface of the star, and because some
of the heat is conducted into the core of the star.  Moreover, while
the temperature peaks in the crust due to individual reactions are
well-separated at the early times, they become blended together by the
time they reach the surface.

Figure \ref{fig:tevol} clearly suggests that, in order to produce
appreciable variability in the thermal emission from the surface, the
energy deposition by the \noneq\ reactions should be sufficient to
heat the crust locally by $\delta T\sim T$.  The typical outburst
fluences indicate that only $\Delta M\lesssim10^{23}$~ grams of
material is accreted during an outburst, and hence
$\lesssim10^{41}$~erg of energy is deposited in the crust.  This
amount is rather small, which allows us to place some interesting
constraints on the heat capacity of NS crust.  The thermal time at the
depths where this energy is deposited is at least comparable to and
usually much greater than a typical outburst duration.  Thus, during
the outburst, the heat is not conducted away, but primarily heats the
crust locally.  Neglecting heat diffusion, the nuclear energy
deposited in the region near neutron drip will heat this region by an
amount
\begin{eqnarray}\label{eq:delta-T-estimate}
\delta T \sim 10^4 {\rm~K~} 
\left(\frac{C}{k_B/{\rm baryon}}\right)^{-1}
\left(\frac{p_d}{10^{30}{\rm~ erg~cm^{-3}}}\right)^{-1}
\\ \nonumber
\times \left(\frac{Q_{\rm nuc}}{1{\rm~MeV}}\right)
\left(\frac{\Delta M}{10^{23} {\rm~g}}\right)
\end{eqnarray}
\noindent where $C$ is the heat capactiy in units of $k_B$ (Boltzmann
constant) per baryon and $p_d$ is the pressure at which the energy is
deposited.  As outlined in \S~\ref{sec:microphysics}, the typical
crustal temperature is $\sim10^{7}-10^{8}$~K.  Therefore, if the heat
capacity of the neutron star crust is $\sim k_B$ per baryon, the crust
relaxation luminosity is only a small perturbation on the overall
cooling of the core.  In order to produce observable variability of
quiescent thermal emission, the NS crust must have $C\ll k_B$ per
baryon. 

\begin{figure}
\begin{center}
\epsfig{file=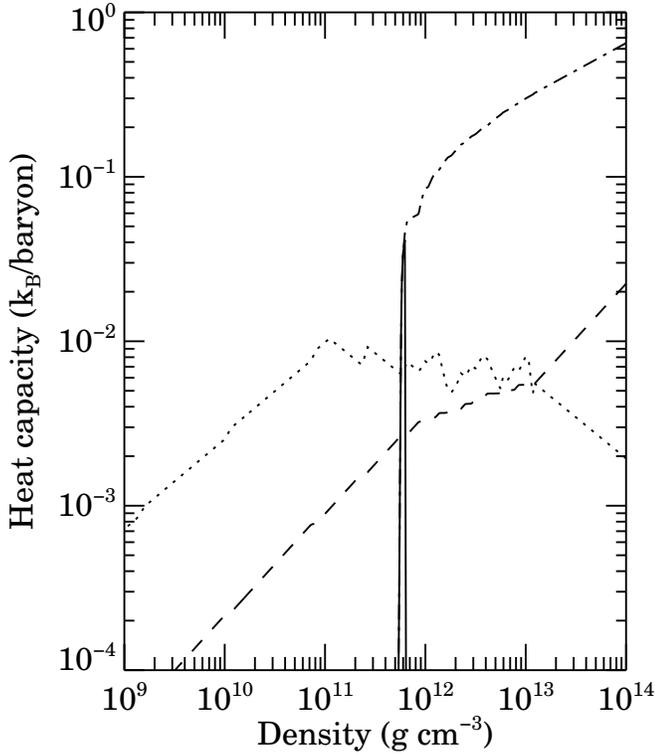}
\end{center}
\caption{ \label{fig:heat-capacity} Heat capacity of neutron star
matter as function of depth, for $\langle \dot{M} \rangle=10^{-11}\
M_\odot$~yr$^{-1}$, in units of $k_B$ per baryon (in these units, the
heat capacity of ideal monoatomic gas would be $C=3/2A$).  Dashed line:
degenerate relativistic electrons, dotted line: ionic lattice
contribution; solid line: degenerate, non-relativistic, superfluid
neutrons; dash-dotted line: degenerate nonrelativistic normal
neutrons. Note that the heat capacity of degenerate neutrons first
rises, and then drops dramatically at the onset of superfluidity, at
$\rho\approx 7\times10^{11}$g~cm$^{-3}$.  The heat-capacities at
$\langle \dot{M} \rangle=$\ee{-10} and \ee{-12} \msun\ \perval{yr}{-1}
are qualitatively similar.  }
\end{figure}

Figure~\ref{fig:heat-capacity} shows contributions to the heat
capacity in the crust, for a representative case of
$\langle\dot{M}\rangle=10^{-11} M_\odot$~yr$^{-1}$.  In the outer
crust, the heat capacity of the ionic latice, $C\lesssim10^{-2} k_B$
per baryon (dotted line, \citenp{shapiro83,vanriper91,chong94}),
dominates.  Strongly degenerate, relativistic electrons (dashed line)
contribute a negligible amount to the heat capacity everywhere except
in the deep inner crust ($\rho\gtrsim2\times10^{13}$~g~cm$^{_3}$).  At
densities above neutron drip, free neutrons, if not superfluid, would
have appreciable heat capacity, approaching $k_B$ per baryon
(dot-dashed line). Thus, if the neutrons in the inner crust are not
superfluid, their heat capacity is so large that the energy release of
$\lesssim10^{41}$~erg characteristic of the heating during an outburst
is not sufficient to heat the crust substantially.  However, it is
commonly believed that the neutrons in the inner crust form a $^1S_0$
superfluid, and their heat capacity is greatly reduced
\cite{maxwell79:_neutr,yakovlev99:_coolin}.  Superfluidity completely
suppresses free neutron heat capacity at densities
$>8\times10^{11}$~g~cm$^{3}$ (solid line in
Figure~\ref{fig:heat-capacity}).  In this case, the ions dominate the
heat capacity in the inner crust, and $C\lesssim10^{-2} k_B$ per
baryon.  Therefore, observing the evolution of quiescent luminosity
provides a test of superfluidity of neutrons: if late-time evolution
is observed ($\delta T/T>$\ee{-3}), then the neutrons must be
superfluid.  While this may not be a controversial conclusion, it is
worth noting that the only existing observational evidence for
superfluidity of crustal neutron gas is the interpretation of pulsar
glitches as being initiated in the crust.  Variability of quiescent
thermal emission from NS transients, if disentangled from accretion in
quiescence or other sources of variability outlined in
\S~\ref{sec:intro}, would be a completely separate direct
observational confirmation of this view.

\begin{figure}
\begin{center}
\epsfig{file=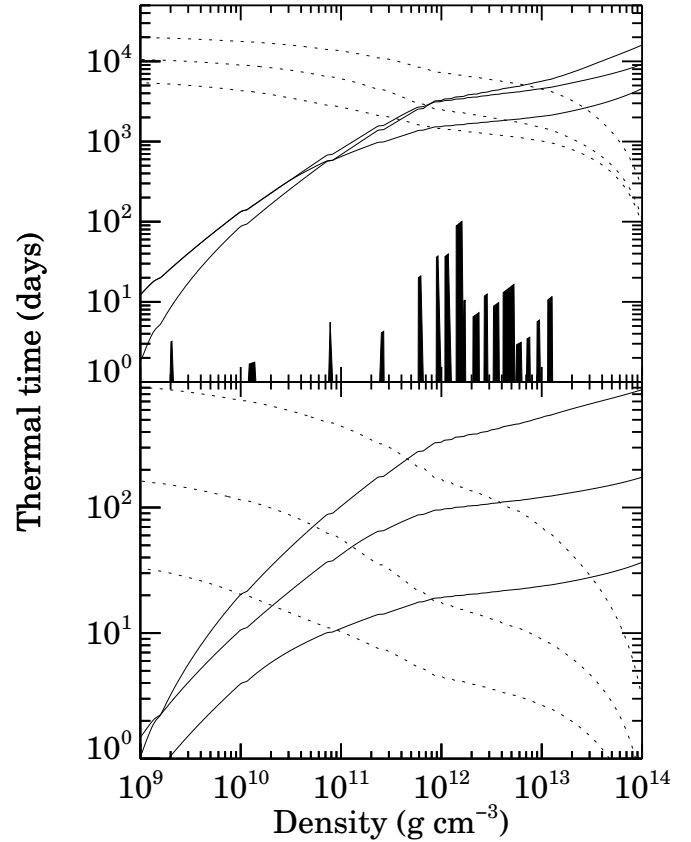}
\end{center}
\caption{ \label{fig:tau} Thermal diffusion time to the surface (solid
lines) and to the core (dotted lines) as function of density for
(reading lines top to bottom) $\langle\dot{M}\rangle=10^{-10}$,
\ee{-11}, and \ee{-12} $\ M_\odot$~yr$^{-1}$.  The top panel
corresponds to models with standard core cooling and low crustal
thermal conductivity, while the bottom panel is for models with
standard core cooling and high crustal thermal conductivity.  The
filled bars in the top panel indicate schematically the regions of
nuclear energy deposition.  The height of the bars is proportional to
the deposited energy, with the highest bar at
$\rho\approx1.5\times10^{12}$~g~cm$^{-3}$ corresponding to 0.47~MeV
per accreted baryon deposited.  }
\end{figure}

The spreading and blending of the temperature peaks due to different
reactions, clearly evident in Figure~\ref{fig:tevol}, as well as the
relative magnitude of heat diffusion towards the surface and the core
can be easily understood in terms of the thermal time at the place
where the heat is deposited.  To make the discussion more precise, we
define the thermal diffusion timescale (\citenp{henyey69}; BBR98)
\begin{equation}
\tau_{\rm th}=\left[\int_{r_1}^{r_2}\left(\frac{\rho
c_v}{K}\right)^{1/2} dr\right]^2.
\end{equation}
Roughly speaking, \tth\ is the time it takes a heat impulse to diffuse
from $r_1$ to $r_2$.  In Figure~\ref{fig:tau}, we plot the thermal
diffusion timescales as a function of density in the crust to the NS
surface (solid line) and the core (dotted lines).  The top panel is
for the model with electron-impurity scattering conductivity in the
limit of very impure crust, while the bottom panel is for the model
with electron-phonon conductivity.\footnote{The thermal times are
shorter for the enhanced cooling models (not shown here) owing to the
lower crustal temperatures.}  Black vertical bars in the top panel
show the locations of the energy deposition due to \noneq\ reactions,
with the height of the bars proportional to the energy deposited.
Clearly, the energy depositions in the inner crust happen at very
similar depths, such that the differences between the thermal times of
the reactions are much smaller than the thermal times themselves.  By
the time the heat from these reactions diffuses to the surface, the
differences between them are blended.  It is also clear from this
figure that if the thermal diffusion time to the core is smaller than
that to the surface, then most of the heat will flow to the core,
rather than to the surface.

\begin{figure*}
\begin{center}
\epsfig{file=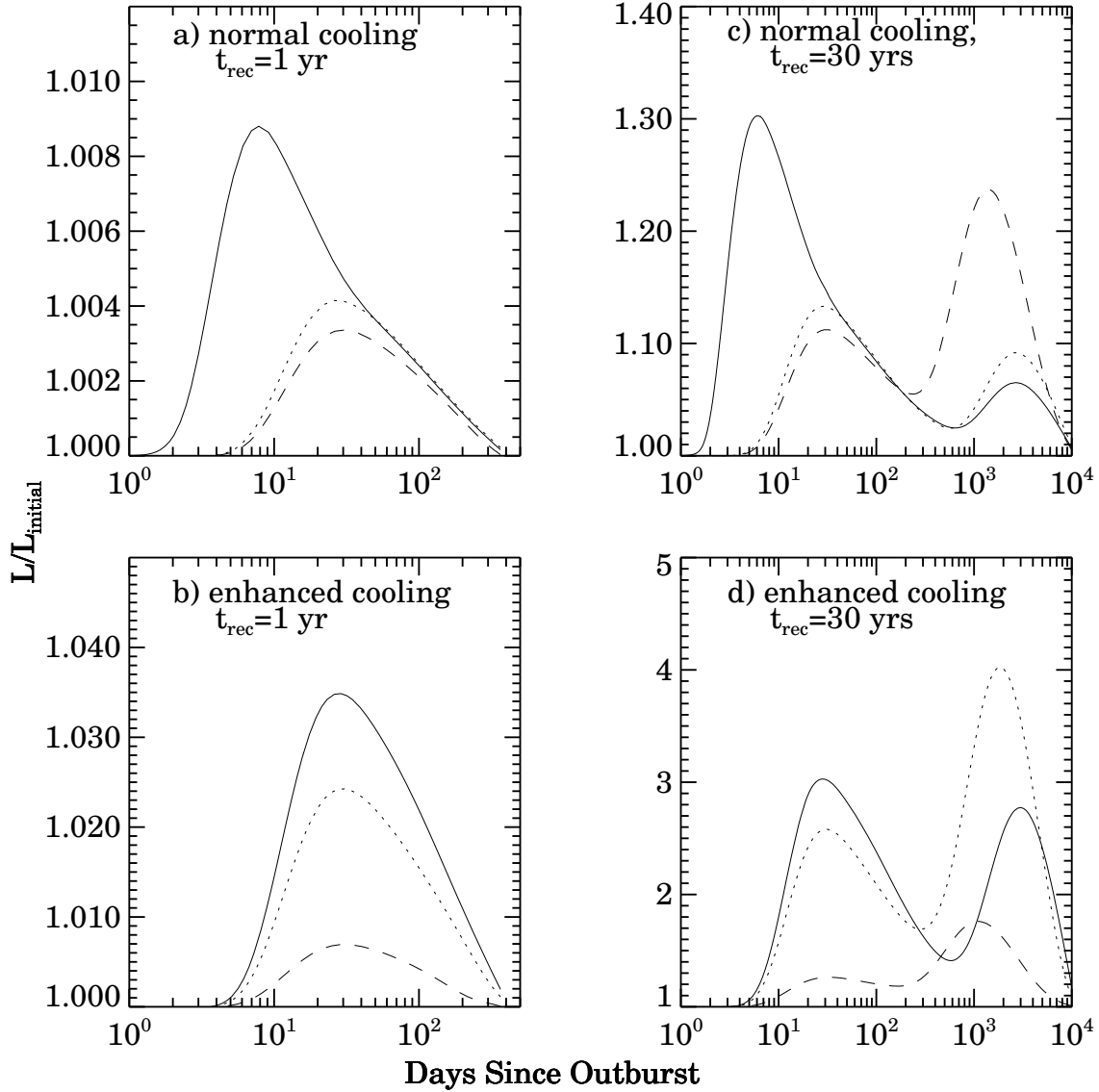}
\end{center}
\caption{ \label{fig:eilc} Thermal emission luminosity from the NS
photosphere, due to crustal heating and NS core temperature
(neglecting accretion luminosity) as a function of time since a
one-day-long ``outburst'' -- for crustal conductivity due to
electron-impurity scattering.  The luminosity is normalized by its
value just prior to the outburst (i.e., by the lowest value over the
recurrence time).  Panels {\em a} and {\em c} are for models with
modified Urca (``standard'') cooling, while panels {\em b} and {\em d}
are for models with accelerated cooling (in this case, pion
condensate). The assumed outburst recurrence time is indicated in each
panel.  The time average accretion rates used are:
$\langle\dot{M}\rangle=10^{-10}$ (solid line), \ee{-11} (dotted line),
\ee{-12} (broken line)$\ M_\odot$~yr$^{-1}$.  }
\end{figure*}

\begin{figure*}
\begin{center}
\epsfig{file=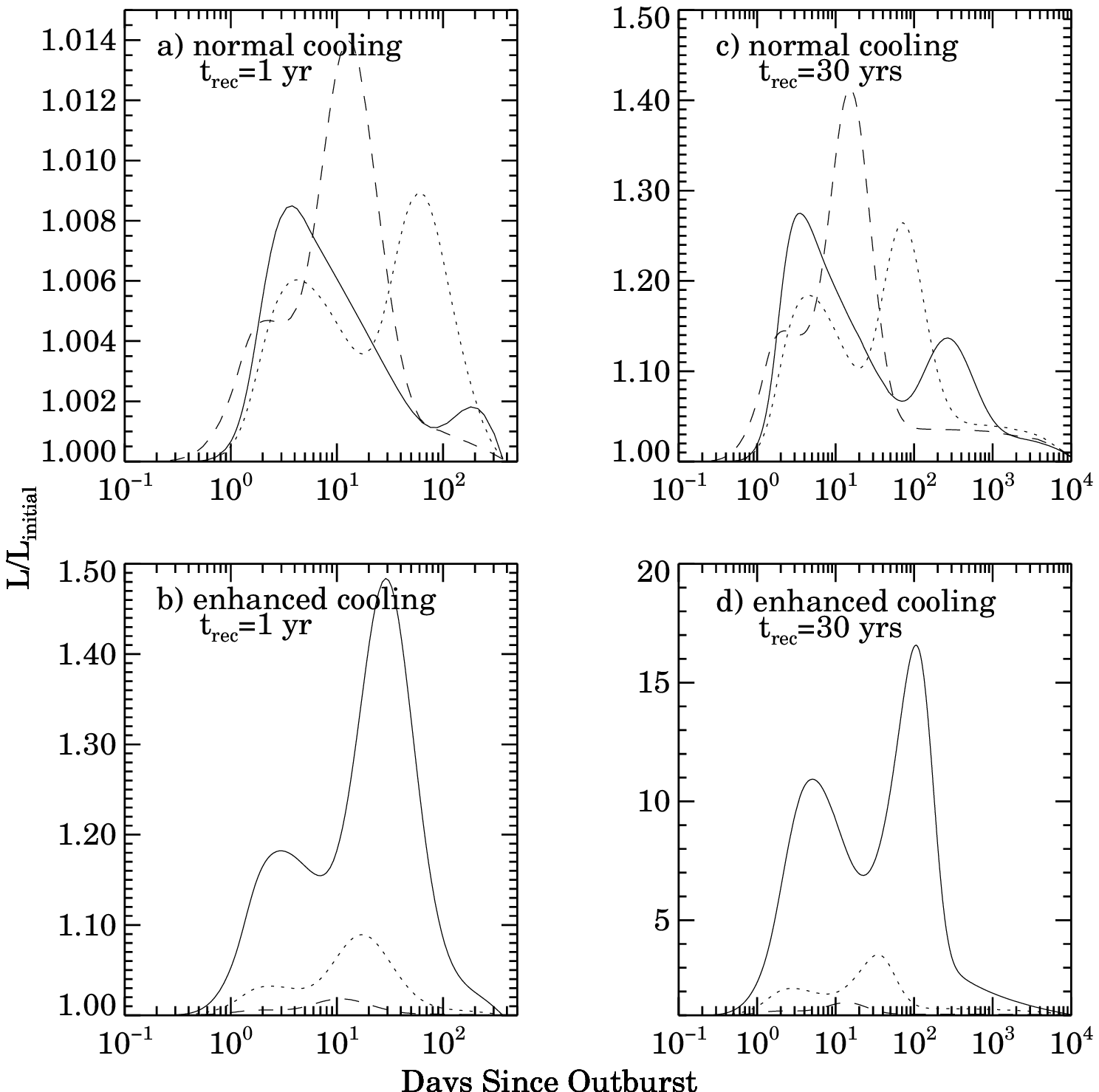}
\end{center}
\caption{ \label{fig:ephlc} Same as Figure~\ref{fig:eilc}, except the
assumed crustal conductivity is due to electron-phonon scattering.  }
\end{figure*}

In Figures~\ref{fig:eilc} and~\ref{fig:ephlc} we survey the dependence
of the transient emission following an accretion outburst on various
model parameters. Figure~\ref{fig:eilc} shows results for a crust with
low thermal conductivity (electron-impurity scattering with $Q_{\rm
imp}=Z^2$), while Figure~\ref{fig:ephlc} shows results for high
thermal conductivity (electron-phonon scattering).  These
figures are plotted with luminosity normalized by the thermal
luminosity just prior to the outburst, which is the lowest luminosity
of the outburst cycle.  This tends to exaggerate visually the range of
luminosities.  It also does not display the absolute differences in
the luminosities due to different \tmdot, which are discussed in
\S~\ref{sec:time-average}.  However, it does make clear the relevant
factors of variability which can be produced over a recurrence cycle
due to this mechanism, and that is our goal here.

As is clear from the figures, the variability is the smallest
($\lesssim1$\%) for models with standard cooling and short (1 year, in
our case) outburst recurrence times.  The variability is largest for
models with rapid core cooling and long (30 year) outburst recurrence
time. Under the same conditions, the variability of the models with
high crustal thermal conductivity (Figure~\ref{fig:ephlc}) is always
larger than that of the low conductivity models
(Figure~\ref{fig:eilc}).  All of these features are straightforward to
understand in light of the above discussion. Since the energy
deposited into the crust is not large ($\lesssim10^{41}$~erg), the
colder the crust, the larger the transient thermal response. Models
with rapid core cooling and high thermal conductivity have
systematically lower crust temperatures (see
\S~\ref{sec:time-average}), and hence have larger amplitude of
variability.

As discussed above, the first hump in the light curve is due to
reactions in the outer crust, while the second hump is due to those in
the inner crust. This is a generic feature of all models except for
the low-$K$, short-recurrence ones (left column of
Figure~\ref{fig:eilc}).  As is evident from the top panel of
Figure~\ref{fig:tau}, the thermal diffusion time from the inner crust
to the surface is $\gtrsim10^3$~days, i.e., longer than the recurrence
time.  Thus, the second peak in the lightcurve does not make it to the
NS surface before the next outburst, and only one peak is observed in
the quiescence lightcurve.  This mismatch between the thermal
diffusion timescale and the outburst recurrence timescale explains why
only one light-curve peak is observed in the \trec=1 yr, low thermal
conductivity simulations (Figures~\ref{fig:eilc}a \& b) while two
peaks are observed in all other simulations.  For long \trec\ sources,
the time between the outburst and the second peak in the lightcurve is
an excellent observational indicator of the thermal time in the inner
crust.

As we see in Figure~\ref{fig:tau}, the thermal time of the inner crust
in low-$K$ models is $\approx10^3-10^4$ days, while it is only
$\sim10^2$ days for the high-$K$ models.  As a result, for low-$K$
models with \trec=1 yr, the crust does not completely cool down, and
therefore remains at a roughly constant temperature.  On the other
hand, for \trec=30 yr, the crust does have enough time to cool down,
and hence the variability is much greater (tens of per cent to factors
of few).  In addition, it is clear from
Eq.~(\ref{eq:delta-T-estimate}) that it is $\Delta
M=\langle\dot{M}\rangle\tau_{\rm rec}$, rather than
$\langle\dot{M}\rangle$ or $\tau_{\rm rec}$ individually that is
important for the overall magnitude of variability.  For a given
$\langle\dot{M}\rangle$, models with larger $\tau_{\rm rec}$ exhibit
larger variability, simply because the amount of energy deposited into
the crust is larger. This explains why the \trec=1 yr light curves are
(relatively) less variable than the \trec=30 yr light curves.  In
general, one therefore expects a greater magnitude of (relative)
intensity variability in quiescence from systems with longer
recurrence timescales (\approxgt 10 yr).

\section{Comparison with Observations}
\label{sec:discuss}

Is the phenomenon we describe here observed in transient NSs in
quiescence?  If so, both the time-averaged luminosity level, and the
variability around it need to be self-consistently accounted for.
This constraint provides an observational discriminant for the various
core cooling scenarios in NS transients.

The observed magnitude of variability (factors of $\sim3-5$) cannot be
achieved in models with standard neutrino cooling, which vary by less
than 50\% in all cases we investigate here.  On the other hand, rapid
neutrino cooling models are capable of producing intensity variability
by factors of up to 20 on timescales of months-years; this may be
invoked for the observed variability of 4U~2129+37 (a factor of 3.4
over months), and the longer time-scale variability of \cenx4\ (factor
of 2-5 over 6 years, an additional factor of 3 over 10 years).  The
factor of 2-3 variability in \aql\ over months-years can be similarly
explained, again requiring rapid cooling to be operative in its core.
However, to explain the factor of $\sim$3 variation over a period of
4-8 days in the quiescent X-ray flux of \cenx4\ observed by Campana
\etal\ \cite*{campana97}, one must invoke an unobserved, short
($\approxlt $ 1 day) outburst, preceeding those observations by only
$\sim$ few days, which seems unlikely, although it cannot be excluded
due to the lack of X-ray coverage.

However, as argued by BBR98, the inferred quiescent luminosity of
\aql\ can be naturally understood if the neutrino losses from its core
are negligible, i.e. if only standard cooling processes operate
(cf. our Eq.~[\ref{eq:Lq-normal-cool}]).  Time-averaged \llqq\ of
\aql\ is {\em inconsistent\/} with rapid cooling, regardless of the
assumed crustal conductivity (cf. Eq.~[\ref{eq:Lq-enh-eph}] and
[\ref{eq:Lq-enh-ei}]).  This is also the case for 4U~1608-52.  For
these sources, variability in quiescence must be due to processes
other than thermal relaxation of the crust (outlined in
\S~\ref{sec:intro}).

Of the observed recurrent NS transients, only \cenx4\ has
time-averaged \llqq\ much lower than given by
Eq.~(\ref{eq:Lq-normal-cool}).  BBR98 conjectured that this
discrepancy is due to the fact that only $Q_{\rm nuc}\sim 0.1$~MeV,
rather than $\sim1$~MeV is deposited in its crust,\footnote{The low
quiescent luminosity could also potentially be explained by an
overestimation of the time-average accretion rate of \cenx4, due to
the small number of outbursts observed.} while Colpi \etal\
\cite*{colpi00b} argued that $Q_{\rm nuc}\sim1$~MeV, but rapid
neutrino cooling is operating in \cenx4.  Clearly, analysis of {\em
variability\/} of \cenx4\ in quiescence (or other transients with long
recurrence times) provides a discriminant between the various core
cooling mechanisms.  If enhanced neutrino cooling is responsible for
the relatevely low \llqq\ of \cenx4, then its quiescent luminosity
must vary in a well-defined way (see panels {\em d\/} of
Figures~\ref{fig:eilc} and \ref{fig:ephlc}).  In the contrary case of
standard cooling with low $Q_{\rm nuc}$, the baseline thermal
luminosity of \cenx4\ should be constant to better than 50\%, and any
variability must be accounted for by the processes outside the NS
crust.  Finally, it is difficult to reliably predict the expected
quiescent luminosity of 4U 2129+47, due to its unusual accretion
history \cite{pietsch86}, and so we draw no additional conclusions for
this source.

If transiently accreting pulsars in quiescence fully exclude accretion
onto the NS surface due to the propeller effect, then observations of
the thermal pulse during quiescence will be due only to crustal and
core emission.  Moreover, optical observations can indicate when the
accretion disk is absent in these sources \cite{finger00}, and
observations which follow the pulsed intensity profile of such an
object over several years following an outburst can readily constrain
the NS atmospheric luminosity. The quiescent emission of transient
pulsars, such as the pulsed emission recently detected from A0535+35
\cite{finger00} and the persistent emission from \saxj1808\
\cite{stella00,dotani00}, can potentially be used for measuring the
late-time emission predicted here.  However, theoretical modeling of
the quiescent emission in this case is complicated by the need to
understand the effects of the magnetic field, transverse heat flow in
the crust, and whether or not the crust is replaced by accretion in
these systems.

\section{Conclusions}
\label{sec:conclude}
We have investigated the time-variable luminosity of a transiently
accreting NS, following a ``delta-function''-like accretion outburst
(one day in duration), for three different values of \tmdot, for NS
cores with both standard and rapid core cooling (specifically, for
$\nu$ emissivity from modified Urca and from pion condensate), and for
high crustal thermal conductivity dominated by electron-phonon
scattering (for the case of a pure crystal crust) or for low
conductivity due to electron-impurity scattering with large impurity
fraction (for the more likely case of very impure crust expected in an
accreting NS).

The magnitude and time-dependence of late-time emission due to heat
deposited by \noneq\ reactions in the crust depends intimately on the
microphysics active in the crust and the core, the time-average
accretion rate (and thus the time-average quiescent luminosity itself)
and -- most importantly -- on the recurrence timescale of the
transient system.  In the case of standard core cooling, the magnitude
of the luminosity swings (peak-to-peak) is \approxlt 1\% for \trec=1
yr, and up to 15-40\% for \trec=30 yr, independent of the crust
conductivity and \tmdot.  For rapid core cooling, the magnitude of the
luminosity swings varies between $\sim$few percent, to as much as a
factor of 20, and is largest for long \trec\ and high crustal thermal
conductivity. 

It is possible to observe either one peak in the lightcurve (due to
the blended emission from the several reactions in the outer crust) or
two peaks (the second is due to blended emission from reactions in the
inner crust).  Observing the second peak requires a recurrence time
long enough to permit the inner-crust emission to reach the surface.
For low crustal thermal conductivity, this thermal time is $\sim10$
yr, so the second peak is only observed in transients with longer
recurrence times.  In the electron-phonon scattering dominated crust,
with thermal time of $\lesssim1$ yr, the second peak may be observed
in short \trec\ sources.  In general, only the energy deposited at
depths where the thermal time is smaller than the outburst recurrence
interval will produce observable variability in the quiescent thermal
emission.

The changes in luminosity which would permit one to follow the
time-variable emission due to deep-crustal heating at the $\sim$few
per cent level are $\delta L_X\sim$\ee{30} \cgslum\ (=5\tee{-16}
$[d/4.0 {\rm kpc}]^2$ \cgsflux), which is detectable with \chandra,
\xmm\ and \conx\ in $\sim$\ee{5}, 5$\times$\ee{4}, and \ee{4} sec,
respectively.  Complicating such observations is the contribution of
accretion onto the NS, which could add a stochastic variability
component to the thermal spectrum which may not easily be
distinguished.  If accretion only occurs at the magnetosphere, this
contributes to the power-law component, which can be separated
spectrally.

Our comparisons between these results and previously reported
observations of transiently accreting NSs in quiescence indicate that
the observed magnitude of intensity variability is much greater than
we predict for standard neutron star core cooling and crust
conductivity (factors of $\sim$few for the former vs. $<50$\% for the
latter).  Hence, for the recurrent transients \aql, 4U~1608-52, and
SAX~J1808-36, whose time-averaged luminosity agrees well with the
predictions of standard neutrino cooling (BBR98; cf. our
\S~\ref{sec:time-average}), it seems likely that some other process --
such as continued, variable accretion in quiescence -- is required to
explain the majority of the observed variability.

In the case of \cenx4, however, the observed quiescent luminosity {\em
is} consistent with rapid core cooling (\citenp{colpi00b}, cf. our
\S~\ref{sec:time-average}), and hence its long-timescale (months to
years) variability could be accounted for by crustal
relaxation.\footnote{But the short timescale of variability ($\sim$few
days) requires that a short outburst (\approxlt days) would have had
to occur only a few days prior to the observation; this seems
unlikely, but cannot be excluded.}  On the other hand, if \cenx4\
appears too dim because its time-averaged accretion rate has been
overestimated or \noneq\ reactions deposit an unusually small amount
of energy into its crust (BBR98), then the majority of its quiescent
variability must be due to stochastic sources external to the NS.
Disentangling the true thermal emission from \cenx4\ from the
contamination due to continued accretion in quiescence (or other
sources) would allow one to discriminate these two hypotheses.

Finally, we note that the observations are low quality (S/N$\sim$few);
there are only two instances of repeated observation with the same
instrument, and in those cases the number of observations were only
2-3 (not densely sampled light curves); and the infered time-average
accretion rate (and, hence, the predicted quiescent luminosity) of Cen
X-4 is highly uncertain due to the small number of outbursts
observed. Therefore, a definitive statment on this subject requires
greater observational scrutiny, with higher S/N data than presently
available, and with densely sampled lightcurve following an outburst,
which would be capable of distinguishing between the temporal
variability predicted here and more stochastic variability which
likely accompanies a variable accretion rate.

RR gratefully acknowledges useful conversations with Ed Brown and Lars
Bildsten, with whom this idea was initially discussed, and whose
comments improved this paper.  The authors also thank Dimitri Yakovlev
for his extensive comments on this paper.  This work was supported in
part by NASA Grants NAG5-3239 and NGC5-7034 and NSF Grant AST-9618537.
GU acknowledges support as a Lee A. DuBridge postdoctoral scholar.

\end{document}